\documentclass[aps,pre,preprint,showkeys,amsmath,amssymb,nofootinbib]{revtex4-1}
\begin{document}
\begin{flushleft}
\textit{Astronomy Letters, 2011, Vol. 37, No. 7, pp. 468--479.}\footnote{%
\baselineskip=5pt
Original Russian Text has been published in Pis'ma v Astronomicheskii Zhurnal, 2011, Vol.~37, No.~7, pp.~512--524.}
\end{flushleft}
\title{Formation of ``Lightnings'' in a Neutron Star Magnetosphere\\
and the Nature of RRATs}
\author{Ya.\ N. \surname{Istomin}}
\email{\texttt{istomin@lpi.ru}}
\author{D. N. \surname{Sobyanin}}
\email{\texttt{sobyanin@lpi.ru}}
\affiliation{Lebedev Physical Institute, Russian Academy of Sciences,\\Leninskii pr.\ 53, Moscow, 119991 Russia}
\received{December 13, 2010}
\begin{abstract}
The connection between the radio emission from ``lightnings'' produced by the absorption of high-energy photons from the cosmic gamma-ray background in a neutron star magnetosphere and radio bursts from rotating radio transients (RRATs) is investigated. The lightning length reaches 1000~km; the lightning radius is 100~m and is comparable to the polar cap radius. If a closed magnetosphere is filled with a dense plasma, then lightnings are efficiently formed only in the region of open magnetic field lines. For the radio emission from a separate lightning to be observed, the polar cap of the neutron star must be directed toward the observer and, at the same time, the lightning must be formed. The maximum burst rate is related to the time of the plasma outflow from the polar cap region. The typical interval between two consecutive bursts is $\sim100$~s. The width of a single radio burst can be determined both by the width of the emission cone formed by the lightning emitting regions at some height above the neutron star surface and by a finite lightning lifetime. The width of the phase distribution for radio bursts from RRATs, along with the integrated pulse width, is determined by the width of the bundle of open magnetic field lines at the formation height of the radio emission. The results obtained are consistent with the currently available data and are indicative of a close connection between RRATs, intermittent pulsars, and extreme nullers.
\end{abstract}
\maketitle

\section*{INTRODUCTION}

Rotating radio transients (RRATs), the sources that manifest themselves in the radio band as separate, sparse, short, relatively bright bursts, were discovered by analyzing archival data from the Multibeam Pulsar Survey with the Parkes 64-m radio telescope and were immediately associated with the manifestation of a bursty behavior of neutron stars (McLaughlin et~al. 2006). The burst rate lies within the range from about $1\text{ min}^{-1}$ to $1\text{ h}^{-1}$. The burst phase is retained in RRATs, which allowed the periods $0.1-6.7$~s slightly exceeding those of normal radio pulsars to be measured (Keane et~al. 2010). The intensity of single radio bursts reaches 310~mJy in observations at 111~MHz (Shitov et~al. 2009) and lies within the range from 100~mJy to 10~Jy in observations at 1.4~GHz (Keane et~al. 2010), considerably exceeding that for normal radio pulsars. Measurements of the period derivative were made for seven RRATs; they confirm the existence of a high surface magnetic field exceeding $2.5\times10^{12}$~G (McLaughlin et~al. 2009). For RRAT J1819--1458, the surface magnetic field reaches $5\times10^{13}$~G (McLaughlin et~al. 2006; Esamdin et~al. 2008) and exceeds the critical one. In this case, the specificity of the electron-positron pair production changes, in particular, due to photon splitting (Chistyakov et~al. 1998; Rumyantsev and Chistyakov 2005), but the plasma generation, along with the accompanying radio emission, is not suppressed even in ultrahigh magnetar magnetic fields (Istomin and Sobyanin 2007, 2008). This is suggested by the observations of magnetars at low (Shitov et~al. 2000; Malofeev et~al. 2005, 2007, 2010) and high (Camilo et~al. 2006, 2007; Levin et~al. 2010) radio frequencies. Nevertheless, the plasma-generated radio emission itself is not necessarily observed constantly and the nonstationary activity of RRATs can be associated with a high magnetic field. For example, the radio emission from the magnetar PSR J1622--4950 with a high surface magnetic field of $2.8\times10^{14}$~G is nonstationary: its flux density changes by a factor of 6 in a day and the radio source itself can switch off for many hundreds of days without having any accompanying X-ray variability (Levin et~al. 2010). Weltevrede et~al. (2010) talk about the existence of additional sporadic components in PSR J1119--6127 with a surface magnetic field of $4.1\times10^{13}$~G that do not manifest themselves in the averaged pulse profile and, in a sense, are RRAT-like.

The radio emission from normal pulsars owes its existence to the outflows of an electron-positron plasma from the magnetosphere (see, e.g., the review by Beskin 1999). However, the bursty radio emission from RRATs is basically nonstationary. Consequently, the regular disappearance of radio emission is indicative of the possible cessation of plasma generation (Gurevich and Istomin 2007), while the generation itself should be nonstationary. This is supported by the measured difference in spindown rate of the intermittent pulsars B1931+24 and J1832+0029 in the periods of ``operation'' and ``silence'' (Lyne 2009). The existence of pulsars with a high nulling fraction reaching 95\% (Wang et~al. 2007) also argues for the possibility of a temporary cessation of plasma generation in the neutron star magnetosphere.

If the plasma generation is off, then efficient filling of a magnetosphere that can initially be in a vacuum state is possible through the absorption of photons from the cosmic gamma-ray background (Istomin and Sobyanin 2009, 2010a, 2010b). Shukre and Radhakrishnan (1982) showed that allowance for the influence of these photons on the particle production processes can be important even when ordinary stationary plasma generation is considered: in the Ruderman-Sutherland model, the influx of new charges to trigger each new spark in the polar gap of a pulsar magnetosphere can be provided by the absorption of photons from the diffuse gamma-ray background.

Istomin and Sobyanin (2011b) showed that the absorption of a high-energy photon from the external cosmic gamma-ray background in the inner neutron star magnetosphere triggers the generation of a secondary electron-positron plasma and produces a lightning --- a lengthening and simultaneously expanding plasma tube. The generation of an electron-positron plasma in a lightning is essentially nonstationary. The multiplicity, i.e., the ratio of the plasma density to the Goldreich-Julian density, exceeds $10^4$, while the dense relativistic electron-positron plasma formed in the short lightning lifetime can manifest itself in the radio band as short single radio bursts. Here, we investigate the possible connection between radio bursts from separate lightnings and those from RRATs.

The paper is structured as follows. In the next section, we investigate the absorption of photons from the cosmic gamma-ray background in a neutron star magnetosphere and estimate the time it takes for the magnetosphere to be filled with plasma. Next, we study the magnetospheric regions where lightnings are efficiently formed and discuss the conditions for observing the radio emission generated by these lightnings. Subsequently, we consider the factors that determine the width of single radio bursts from RRATs and investigate the possible mechanism explaining the observed spread of radio bursts in phase. The rate of radio bursts from RRATs is then estimated. In conclusion, we present our main results.

\section*{THE ABSORPTION OF GAMMA-RAY PHOTONS AND THE FILLING OF A MAGNETOSPHERE WITH PLASMA}

Consider a neutron star with a magnetosphere in a vacuum state. This implies that the plasma density in the magnetosphere should be low compared to the Goldreich-Julian density $\rho_{GJ}\approx-\mathbf{\Omega}\cdot\mathbf{B}/2\pi c$ (Goldreich and Julian 1969), where $\mathbf{\Omega}$ is the angular velocity of the neutron star, $\mathbf{B}$ is the magnetic field, and $c$ is the speed of light. The electromagnetic field for this case was calculated by Deutsch (1955). There is a high longitudinal electric field $E_\parallel$ in such a magnetosphere whose characteristic strength is defined by the relation $E_\parallel/B\sim\Omega R_S/c=R_S/R_L\sim10^{-4}$, where $R_S$ is the neutron star radius, $R_L=c/\Omega=cP/2\pi$ is the light cylinder radius, and $P$ is the neutron star rotation period. Let us calculate the time it takes for the neutron star magnetosphere to be filled with a dense electron-positron plasma. If there is no free charge escape from the neutron star surface, then the magnetosphere will be filled through the absorption of photons from the external cosmic gamma-ray background (Istomin and Sobyanin 2009, 2010a, 2010b).

Below, we will work in a dimensionless system of
units by measuring the electric and magnetic field
strengths in units of the critical field
\begin{equation*}
\label{criticalField}
B_{cr}=\frac{m_e^2c^3}{e\hbar}\approx4.414\times10^{13}\text{ G},
\end{equation*}
where $m_e$ is the electron mass, $e$ is the positron charge, and $\hbar$ is the Planck constant. As the units of mass, length, and time, we will take the electron mass $m_e\approx9.109\times10^{-28}$~g, the Compton electron wavelength $^-\!\!\!\!\lambda=\hbar/m_e c\approx3.862\times10^{-11}$~cm, and its ratio to the speed of light $^-\!\!\!\!\lambda/c\approx1.288\times10^{-21}$~s, respectively. Thus, the length, velocity, time, and energy will be measured in units of~$^-\!\!\!\!\lambda$, $c$, $^-\!\!\!\!\lambda/c$, and~$m_e c^2$, respectively. All quantities, unless their dimensions are given explicitly, are assumed to be written in dimensionless units. Note the necessary relations $1\text{ cm}\approx2.590\times10^{10}$ and $1\text{ s}\approx7.763\times10^{20}$.

Let a photon have an energy $k$ and propagate in a magnetic field $B$ at an angle $\chi$ to the magnetic field direction. When the transverse photon momentum component exceeds~2, the one-photon electron-positron pair production process becomes possible. The photon absorption coefficient in a magnetic field was calculated by Klepikov (1954) and is
\begin{equation}
\label{weakFieldPhotonAttenuationCoefficient}
\kappa(B,k,\chi)=b\alpha B\sin\chi\exp\Big(-\frac{8}{3}\frac{1}{Bk\sin\chi}\Big),
\end{equation}
where $b=3\sqrt{3}/16\sqrt{2}\approx0.23$, $\alpha=e^2/\hbar c\approx1/137$ is the fine-structure constant. We will take the photon absorption coefficient in a magnetic field in the form
\begin{equation}
\label{simplifiedAbsorbtionCoefficient}
\kappa(B,k)=b\alpha B\exp\Big(-\frac{8}{3Bk}\Big),
\end{equation}
where, in comparison with Eq.~\eqref{weakFieldPhotonAttenuationCoefficient}, there is no dependence on the angle~$\chi$. Here, we use the fact that the photons coming into the magnetosphere from outside propagate mostly at relatively large angles to the magnetic field. To simplify our calculations, we will also take the magnetic field of the neutron star itself to be spherically symmetric:
\begin{equation*}
\label{sphericallySymmetricMagneticField}
B(r)=B_0\Bigl(\frac{R_S}{r}\Bigr)^3,
\end{equation*}
where $B_0$ is the surface magnetic field of the neutron star and $r$ is the distance from its center. A magnetosphere with a spherically symmetric photon absorption coefficient will then be associated with the neutron star. To reach the conclusions needed to obtain our main results, it will suffice to have only a qualitative estimate for the absorption rate of Galactic photons in the neutron star magnetosphere. Therefore, these simplifications are justified.

Let us specify the photon energy $k$ and find the absorption cross section for photons with this energy in the magnetosphere. For this purpose, it is necessary to fix some impact parameter $r_\sigma$ and to specify the photon path by the equation
\begin{equation}
\label{photonLine}
r(x)=\sqrt{r^2_\sigma+x^2},\qquad-\infty<x<\infty,
\end{equation}
where $x$ is the length along the path measured from the point $r=r_\sigma$ nearest to the neutron star. The impact parameter $r_\sigma$ will then be determined from the condition for the integral of the photon absorption coefficient \eqref{simplifiedAbsorbtionCoefficient} along the photon path \eqref{photonLine} being equal to unity:
\begin{equation*}
\label{rSigmaCondition}
\int\limits_{-\infty}^\infty\kappa\bigl(B(r(x)),k\bigr)\,dx=1.
\end{equation*}

Let us introduce a magnetic field $B_\sigma= B(r_\sigma)$ corresponding to the point $x=0$ and a quantity
\begin{equation*}
\label{lambdaSigma}
\Lambda_\sigma=\frac{8}{3B_\sigma k}.
\end{equation*}
It should be assumed that $\Lambda_\sigma\gg1$. This condition is a criterion for the applicability of Eq.~\eqref{simplifiedAbsorbtionCoefficient} for the magnetic field $B=B_\sigma$, which allows this formula to be used at any point of the photon path~\eqref{photonLine}. We can obtain
\begin{equation*}
\label{finalLambdaSigma}
\Lambda_\sigma=\Lambda^{(0)}_\sigma-\frac{7}{6}\ln\Lambda_\sigma,
\end{equation*}
where
\begin{equation*}
\label{lambdaSigma0}
\Lambda^{(0)}_\sigma=\ln\biggl(\sqrt{\frac{2\pi}{3}}\,b\alpha R_S B_0^{1/3}\Bigl(\frac{8}{3k}\Bigr)^{2/3}\biggr).
\end{equation*}
At large $\Lambda^{(0)}_\sigma$, we can write out an approximate solution:
\begin{equation*}
\label{approxLambdaSigma}
\Lambda_\sigma\approx\Lambda^{(0)}_\sigma-\frac{7}{6}\ln\Lambda^{(0)}_\sigma.
\end{equation*}
For a neutron star radius $R_S\approx2.6\times10^{16}$ (10~km in dimensional units), a surface magnetic field $B_0\sim0.01-0.1$, and a gamma-ray photon energy from a wide range $k\sim2-10^7$, we obtain $\Lambda_\sigma\sim20-30$. As was assumed, $\Lambda_\sigma\gg1$. The cross section of the neutron star magnetosphere for photons with energy $k$ is then
\begin{equation*}
\label{magnetosphericCrossSection}
\sigma_{m}(k)=\pi\,r^2_\sigma(k),
\end{equation*}
where
\begin{equation*}
\label{finalRsigma}
r_\sigma(k)= R_S\Bigl(\frac{\Lambda_\sigma}{a_0}\frac{k}{2}\Bigr)^{1/3}
\end{equation*}
and $a_0=4/3B_0$.

Let the flux of photons from the external cosmic gamma-ray background be specified by some function $j_{GB}(k)$, so that the number of photons $dN_{GB}$ that pass in a time $dt$ through a surface $dS$ almost perpendicular to it and have the wave vectors lying in a solid angle $d\Omega$ and the energies lying within a range $dk$ around $k$ is
\begin{equation*}
\label{dNgb}
dN_{GB}=j_{GB}(k)\,dS\,d\Omega\,dk\,dt.
\end{equation*}
We assume $j_{GB}(k)$ to be some effective function corresponding to an isotropic distribution of photons. The number of photons absorbed per unit time in the magnetosphere and having an energy $k$ exceeding some value of $k_{\min}$ is then
\begin{equation}
\label{absorbedFlux}
F^{abs}_{GB}(k_{\min})=4\pi\int\limits_{k_{\min}}^\infty j_{GB}(k)\sigma_{m}(k)\,dk.
\end{equation}

We will take the function $j_{GB}(k)$ as a power law:
\begin{equation}
\label{powerJgb}
j_{GB}(k)=A_{GB}k^{-\beta},
\end{equation}
where $\beta$ is the spectral index. In any case, $\beta>2$. The constants $A_{GB}$ and $\beta$ can be found from experiments. The experimental data usually give $\beta$ and the total particle flux through a unit surface area and in a unit solid angle, with the particles having an energy $k$ exceeding some value of $k^{exp}_{\min}$. We will designate this flux as
\begin{equation*}
\label{experimentalGBflux}
I^{exp}_{GB}=\int\limits_{k^{exp}_{\min}}^\infty j_{GB}(k)\,dk.
\end{equation*}
Then,
\begin{equation*}
\label{Agb}
A_{GB}=I^{exp}_{GB}(\beta-1)(k^{exp}_{\min})^{\beta-1}.
\end{equation*}
Integrating dependence \eqref{powerJgb} yields
\begin{equation}
\label{absFluxViaIexpGB}
F^{abs}_{GB}(k_{\min})=4\pi I^{exp}_{GB}\,\sigma^{exp}_{\min}\,
\frac{\beta-1}{\beta-5/3}
\left(\frac{k_{\min}}{k^{exp}_{\min}}\right)^{-\beta+5/3}.
\end{equation}
Here, we introduced the designation $\sigma^{exp}_{\min}=\sigma_{m}(k^{exp}_{\min})$ for the magnetospheric cross section corresponding to the photon energy~$k^{exp}_{\min}$.

Istomin and Sobyanin (2011b) showed that the absorption of a high-energy photon from the cosmic gamma-ray background in the inner neutron star magnetosphere gives rise to a lightning. The number of electron-positron pairs produced in a lightning through the absorption of one photon reaches~$10^{28}$. Consequently, the filling of the magnetosphere must be very efficient. However, not every photon can give rise to a lightning with such a large number of particles. Let $r_0$ be the distance from the neutron star center at which the primary electron-positron pair that triggers a lightning is produced. When calculating the number of particles, we required that the primary photon be absorbed at a distance $r_0$ much greater than the neutron star radius~$R_S$. If these distances become comparable, then the number of produced particles will decrease catastrophically due to a sharp shortening of the plasma tube. The photon that triggered such a lightning then cannot be taken into account in Eq.~\eqref{absorbedFlux}, because it will not contribute significantly to the total number of particles accumulating in the neutron star magnetosphere compared to other photons. Obviously, $r_0\leqslant r_\sigma(k)$. This consideration allows $k_{\min}$ to be determined from the condition
\begin{equation*}
\label{kMinCond}
r_\sigma(k_{\min})=R_S,
\end{equation*}
which leads us to the formula
\begin{equation}
\label{kMinValue}
k_{\min}=\frac{2a_0}{\Lambda_\sigma}.
\end{equation}

To estimate the time it takes for the magnetosphere to be filled, let us take the currently available Fermi Large Area Telescope data on the cosmic gamma-ray background spectrum (Abdo et~al. 2010). Consider the photons from the extragalactic isotropic gamma-ray background. This will allow us to obtain an upper limit for the time it takes for the neutron star magnetosphere to be filled, because the total gamma-ray background is the sum of the diffuse Galactic component and the isotropic extragalactic component, with the latter making a smaller contribution to the total photon flux than the former. Let us take $\beta=2.41\pm0.05$, $k^{exp}_{\min}=200$ and $I^{exp}_{EGB}=(1.03\pm0.17)\times10^{-5}\text{ cm}^{-2}\text{s}^{-1}\text{sr}^{-1}$. We will also assume that the spectrum can be extended to the range of low energies with the same index $\beta$. In addition, we will take $a_0/\Lambda_\sigma\sim10$.  Using Eqs.~\eqref{absFluxViaIexpGB} and~\eqref{kMinValue}, we can then obtain the characteristic flux $F^{abs}_{EGB}(k_{\min})\sim2\times10^{10}\text{ s}^{-1}$ of photons from the extragalactic background being absorbed per unit time in the neutron star magnetosphere. Istomin and Sobyanin (2011b) showed that the absorption of 10--100 photons is sufficient for the inner neutron star magnetosphere to be filled. We see that the calculated flux of absorbed photons is more than enough for the necessary number of lightnings to be formed to fill the inner neutron star magnetosphere. Consequently, not the flux of photons absorbed in the magnetosphere but the lightning formation time, which determines the magnetosphere filling time, comes to the fore.

Obviously, at great distances from the neutron star surface, the plasma generation efficiency will fall, because both electric and magnetic field strengths decrease. In particular, this is also related to the increase in particle acceleration time causing a reduction of the source of electron-positron pairs (Istomin and Sobyanin 2011a). Recall that an electron or positron moving along a magnetic field line with a radius of curvature $\rho$ is accelerated in a longitudinal electric field $E_\parallel$ in a time $\tau_{st}=\gamma_0/E_\parallel$ to the stationary Lorentz factor
\begin{equation*}
\label{gammaMax}
\gamma_0=\left(\frac{3}{2\alpha}E_\parallel\rho^2\right)^{1/4}
\end{equation*}
in the absence of screening of the electric field reaching $10^8$ (Istomin and Sobyanin 2009, 2010a). There exists some threshold distance $R^{eff}$ at which the formation of lightnings is completely absent. It can be determined by equating the particle acceleration time $\left.\tau_{st}\right|_{r=R^{eff}}$ to the distance $R^{eff}$:
\begin{equation}
\label{Reff}
R^{eff}\sim R_S\biggl(\frac{R_S}{\tau_{st}}\biggr)^{2/5}.
\end{equation}
When calculating $\left.\tau_{st}\right|_{r=R^{eff}}$, we took the vacuum electric field from Deutsch (1955) at distance $R^{eff}$ from the neutron star center and assumed that $\rho\sim R^{eff}$. In Eq.~\eqref{Reff}, by $\tau_{st}$ we mean the time of total particle acceleration near the neutron star surface. Taking a neutron star radius $R_S\approx10$~km and an electric field $E_0\sim10^{-4}$ near the neutron star surface, we obtain $R^{eff}/R_S\sim70$, i.e., the distance $R^{eff}$ can exceed the neutron star radius by one or two orders of magnitude, reaching $\sim1000$~km. Note that $R^{eff}$ is comparable in order of magnitude to the distance derived from the condition for the reduction in photon absorption efficiency due to a decrease in the magnetospheric magnetic field as one recedes from the neutron star (Istomin and Sobyanin 2011b).

Thus, the time it takes for the inner magnetosphere to be filled is limited from below by $R_S/c$ and can reach values of the order of $R^{eff}/c$. The entire magnetosphere will be filled in the time of the plasma transition from the inner regions to the regions near the light cylinder, $R_L/c=P/2\pi$.

\section*{THE FORMATION OF LIGHTNINGS}

We see that the neutron star magnetosphere can be efficiently filled with an electron-positron plasma. It is important to note that here we are dealing with a closed magnetosphere. Nevertheless, open magnetic field lines emerging from the polar cap regions and going beyond the light cylinder to infinity exist in the magnetosphere. The electron-positron plasma can freely escape from the magnetosphere along open magnetic field lines, with the characteristic plasma outflow time being~$R_L/c$.

Let there be a neutron star in which the inner closed magnetosphere is filled with an electron-positron plasma, while the polar cap regions contain no plasma. The photons from the external cosmic gamma-ray background will be efficiently absorbed in both closed and open magnetospheres. Here, three cases can be distinguished. In the first case, a gamma-ray photon can be absorbed in the inner magnetosphere filled with a dense plasma. It will produce an electron-positron pair, but no cascade generation of an electron-positron plasma will subsequently be triggered due to the absence of efficient particle acceleration. In the magnetospheric regions under consideration, the longitudinal electric field is zero, while the transverse electric field does not lead to particle acceleration but only ensures the fulfilment of the corotation conditions. In the second case, a gamma-ray photon can be absorbed in the outer magnetosphere at distances $r>R^{eff}$. In principle, a dense plasma can be absent in these regions. The longitudinal electric field strength is much smaller than the field strength near the neutron star surface but, nevertheless, is nonzero. The absorption of a gamma-ray photon can then give rise to a lightning, but the number of particles in it will be small compared to the number of particles in lightnings generated near the stellar surface in a vacuum electromagnetic field. In the third case, a gamma-ray photon is absorbed in the polar cap regions of the neutron star magnetosphere. We will then have a lightning in which the number of particles is more than enough for the polar cap region to be filled. The formed dense electron-positron plasma will screen the longitudinal electric field as soon as the lightning reaches the neutron star surface. Subsequently, the absorption of gamma-ray photons in the polar cap region will not trigger a generation cascade for the same reasons as those in the first case. A high longitudinal electric field close to the vacuum one is needed for the formation of a new lightning (Istomin and Sobyanin 2011b). This electric field can emerge only after the outflow of the electron-positron plasma produced by a preceding lightning from the polar cap regions of the magnetosphere. Hence we obtain a lower limit on the formation period $\Delta T$ of two consecutive lightnings:
\begin{equation}
\label{lightningPeriod}
\Delta T\gtrsim R_L.
\end{equation}
Here, we took the time of the total plasma outflow beyond the light cylinder as the sought-for time. In principle, this estimate can be reduced if the plasma outflow from the regions where the formation of lightnings is fairly efficient into the regions where it no longer takes place is assumed to be sufficient for the formation of another lightning. In this case, some effective parameter $R^{eff}_L<R_L$ that can reach values of the order of $R^{eff}$ can be substituted into Eq.~\eqref{lightningPeriod} for $R_L$:
\begin{equation*}
\label{ReffL}
R^{eff}<R^{eff}_L<R_L.
\end{equation*}
Formula \eqref{lightningPeriod} can then be rewritten as
\begin{equation}
\label{alternativeLightningPeriod}
\Delta T\gtrsim R^{eff}_L.
\end{equation}

Having investigated the parameters of the generated electron-positron plasma, Istomin and Sobyanin (2011b) showed that if a neutron star located at a distance of the order of the distance to a typical pulsar is considered, then the possibility of observing radio bursts from lightnings at existing radio astronomy observatories can be assumed. Moreover, one may expect the radio flux density from a lightning to exceed that from normal radio pulsars due to larger particle Lorentz factor, multiplicity, and plasma density. To be able to observe the radio emission from a separate lightning, it is necessary that, first, the polar cap of the neutron star be directed toward the observer at some instant of time and, second, the lightning be formed at the same time. For the first condition to be met, the line of sight must lie within the emission cone. The emission cone itself can be determined by several factors. Clearly, the emission cone definitely lies within the encompassing cone formed by the tangents to the last closed magnetic field lines in the polar cap region of the neutron star magnetosphere. Significantly, the opening of the encompassing cone depends on the height at which radio emission at some fixed frequency is generated. In general, we do not know this height a priori. Nevertheless, we can introduce some effective parameter $z^{eff}$ to characterize the opening of the encompassing cone. Introducing the angle $\theta^{\max}_c$ between the generatrix of this cone and the direction of the magnetic axis $\mathbf{m}$ and finding the tangents to the last closed magnetic field lines at distance $z^{eff}$ from the neutron star center, we have
\begin{equation*}
\label{thetaCmax}
\theta^{\max}_c=\frac{3}{2}\sqrt{\frac{z^{eff}}{R_L}}.
\end{equation*}
We will note at once that here we do not consider the influence of the relative positions of the neutron star rotation axis $\mathbf{\Omega}$ and its magnetic axis $\mathbf{m}$ on the width and shape of the emission cone. The possible magnetic field nondipolarity can have a certain influence on the emission cone shape if the emission itself is formed near the neutron star surface. Incidentally, the nondipolarity can be significant for normal pulsars when considering the generation of an electron-positron plasma in the polar gap and when calculating the characteristics of the curvature radiation being formed (Barsukov et~al. 2009).

Obviously, $z^{eff}\geqslant R_S$. Since the parameter $z^{eff}$ takes on different values for different lightnings, the angle $\theta^{\max}_c$ is not strictly fixed. It can be asserted that $z^{eff}$ can exceed the neutron star radius by the length of the forming plasma tube. The order-of-magnitude distances from the stellar center at which lightnings can still be efficiently formed are specified by the quantity $R^{eff}$, which allow us to write
\begin{equation}
\label{zEffEstimate}
R_S<z^{eff}<R^{eff}.
\end{equation}
If the radio emission is formed at large heights $z^{eff}\gg R_S$, then the nondipolarity of the magnetospheric magnetic field may be disregarded. Radio polarization measurements for RRAT J1819--1458 (Karastergiou et~al. 2009) are indicative of a dipolar magnetic field structure in the radio emission zone. If the nondipolar magnetic field components near the neutron star surface are assumed to be significant, then the described dipolar structure can indirectly testify to the formation of radio emission at large heights.

The angular radius $\theta_c$ of the emission cone is smaller than that of the encompassing cone, $\theta_c\leqslant\theta^{\max}_c$. However, in general, the lightning radius $R$ is comparable to the polar cap radius (Istomin and Sobyanin 2011b):
\begin{equation}
\label{radiusEstimation}
R\lesssim100\text{ m}.
\end{equation}
This implies that the angular radius of the emission cone is comparable to that of the corresponding encompassing cone:
\begin{equation}
\label{thetaCsinThetaCmax}
\theta_c\sim\theta^{\max}_c.
\end{equation}

Yet another factor determining $\theta_c$ is the specific instant of time at which the lightning is observed, namely the time interval between the production of a primary electron-positron pair and the time of observation. If this interval is small, then the plasma tube radius is small compared to the radius calculated from Eq.~\eqref{radiusEstimation}. Hence it follows that we can give only an upper limit on the angular radius $\theta_c$ of each specific plasma tube:
\begin{equation*}
\label{thetaCineq}
\theta_c\lesssim\theta^{\max}_c.
\end{equation*}
Nevertheless, relation \eqref{thetaCsinThetaCmax} can be used for characteristic estimation.

\section*{THE WIDTH OF RADIO BURSTS AND THE PHASE DISTRIBUTION}

Let us introduce the window of radiation
\begin{equation*}
\label{deltaP}
\Delta P=\frac{\theta^{\max}_c}{\pi}P,
\end{equation*}
which is the maximum time interval during which the radiation from the polar cap region forming at distance $z^{eff}$ from the neutron star center can potentially be observed. Obviously, a lightning can be observed during a shorter time interval. First, the line of sight may not pass near the center of the emission cone. Second, the plasma tube radius may be smaller than the polar cap radius.

We will disregard the influence of radio signal propagation through the interstellar medium on the observed pulse width. The width of the observed pulse corresponding to a radio burst then cannot exceed $\Delta P$. For the first 11 discovered RRATs in observations at $1.4$~GHz, it is $2-30$~ms (McLaughlin et~al. 2006). To be specific, $\Delta P$ can be estimated from the formula
\begin{equation}
\label{deltaPsim}
\Delta P=3\sqrt\frac{z^{eff}P}{2\pi c},
\end{equation}
in which the dimensions of the quantities were restored. If we take $P\sim1$~s and $z^{eff}\sim R_S\approx10$~km, then $\Delta P\sim7$~ms. If, alternatively, we set $z^{eff}=100R_S$, then $\Delta P$ will increase by an order of magnitude. If, in addition, we take into account the dependence on the period for the same 11~RRATs lying within the range $0.4-7$~s and the fact that the lightning width can be smaller than the polar cap width, then the observed pulse width, in general, can be explained if it is assumed to be determined by the width of the forming plasma tube.

As observations show, the pulse width for the same RRAT differs from pulse to pulse. For example, according to Keane et~al. (2010) most of the bursts from RRAT J1841--14 have a typical width of about 2~ms, but the width of several recorded bursts reaches 20~ms. It follows from Eq.~\eqref{deltaPsim} at $z^{eff}\approx10$~km and $P\approx6.598$~s that the time interval corresponding to the polar cap width for RRAT J1841--14 is 17.8~ms. This value is very close to the maximum observed burst width.

However, the lightning width determines the observed pulse width only if the lifetime of a given tube is longer than the time of its passage through the line of sight. In dimensionless units, the lifetime of a lightning is equal in order of magnitude to the distance from the point of primary electron-positron pair production to the neutron star surface. Once the dimensions have been restored, we can then write
\begin{equation}
\label{lightningLifetime}
T_{l}\sim\frac{r_0-R_S}{c}.
\end{equation}
At $r_0\sim100R_S$, we have $T_l\sim3.3$~ms. This explains the presence of short radio bursts whose duration is much less than the corresponding polar cap width.

Let us turn to the timing data for RRAT J1819--1458. Esamdin et~al. (2008) showed that the arrival times of individual radio pulses have a spread in phase: the bursts are distributed in the time interval from $-48.57$ to $39.71$~ms relative to zero phase corresponding to the exact rotation period, forming a bimodal distribution. Lyne et~al. (2009) provided more complete data from which it follows that the distribution of radio bursts in pulse arrival times actually has a three-component structure. This distribution contains the main central component into which about $60$\% of bursts fall and two side components shifted relative to the central one by $\pm45$~ms. In this case, the full width of the range within which the pulse arrival times lie is $120$~ms ($\pm60$~ms relative to zero phase). The width of a single pulse is much smaller, being $3$~ms at 1.4~GHz. The addition of the profiles for individual radio bursts yields a wide triple-humped averaged profile similar to the pulse profiles observed for normal radio pulsars (Lyne and Manchester 1988).

Note that the pulse arrival times for other RRATs also change from pulse to pulse and lie within some window whose typical width is several percent of the rotation period. According to McLaughlin et~al. (2009), the rms deviation of the pulse arrival times at 1.4~GHz lies within the range from 0.8~ms for RRAT J1826--1419 ($P\approx0.771$~s) to 11.2~ms for RRAT J0847--4316 ($P\approx5.977$~s) and essentially coincides with the width of the averaged profiles obtained by adding the profiles of single pulses. For the RRATs investigated in the paper, no evidence was obtained for the existence of a multicomponent structure in the averaged profile similar to the structure of the pulse profile for RRAT J1819--1458 and all averaged profiles and single-component ones.

If we associate single radio bursts from RRATs with the radio emission from separate lightnings, then the full width of the time interval will be determined by the width of the polar cap region or, more precisely, by the angular radius $\theta^{\max}_c$ of the encompassing emission cone. In particular, the phase distribution of bursts will be determined by the point at which the primary electron-positron pair is produced, by the length that the forming lightning can reach, and by the instant of time and the regions in which the line of sight will cross the plasma tube. In the case of a plasma-filled closed magnetosphere, all lightnings will appear only in the region of open magnetic field lines. The phase of a specific single radio pulse will be determined by the distance from the plane passing through the magnetic and rotation axes to the formation region of radio emission at some fixed frequency at which the observations are performed located in the plasma tube. This distance is a random variable that is definitely limited from above by the radius of the section of the region of open magnetic field lines by the plane orthogonal to the magnetic axis and located at distance $z^{eff}$ from the neutron star center. Taking $P\approx4.263$~s and $\Delta P\approx120$~ms for RRAT J1819--1458 from observational data (Lyne et~al. 2009), we can obtain the effective parameter $z^{eff}\approx700$~km using Eq.~\eqref{deltaPsim}, which is comparable to the estimates of $R^{eff}$ \eqref{Reff}. If the emission is assumed to be formed at such a height, then the lightning length should reach lengths of the same order of magnitude. Based on Eq.~\eqref{lightningLifetime}, we obtain a typical lightning lifetime $T_l\sim2.3$~ms, which is close to the observed width of a single pulse, 3~ms.

Most of the radio bursts from RRAT J1819--1458 are single-component ones. However, 5 of the 162 recorded radio bursts have a two-component structure, with the time interval between the individual components lying within the range from 6 to 16.5~ms (Esamdin et~al. 2008). Note that here we are dealing with single radio bursts rather than the averaged profile obtained by adding the profiles of individual radio bursts. Since two-component pulse profiles are also encountered in normal pulsars, it can be assumed that a stationary emission cone is formed in the time of burst observation (Radhakrishnan and Cooke 1969). This can be an argument for the fact that, at least for some bursts, the lifetime of the lightnings emitting them exceeds the time of passage of the line of sight through the emission cone formed by a separate lightning. In this case, the plasma tube width should be comparable to the polar cap width, which also follows from the analogy with normal pulsars. Formula~\eqref{radiusEstimation} shows that the plasma tubes with the required width are actually formed. Besides, calculations similar to those performed above for RRAT J1819--1458 with a period $P\approx4.263$~s give a polar cap width of 14.3~ms. This value is very close to the observed maximum separation between the components in a two-component radio burst, 16.5~ms. Obviously, the lightning lifetime should not be less than this value. We then obtain a formal lightning length $\sim500R_S\approx5000$~km that exceeds $R^{eff}$. This suggests that either a short-term switch-on of the inner gap is possible once some of the lightnings have reached the neutron star surface or the electron-positron plasma produced in the relatively short lightning lifetime will generate radio emission already during its outflow along open magnetic field lines.

Let us briefly discuss these two possibilities. The switch-on of the inner gap in RRAT can occur, for example, because of neutron star surface heating. It is provided by a short-term passage of the electric current through the polar cap during a stroke of lightning. Note that when the lightning reaches the neutron star surface, the electric current reaches huge values of $10^{13}-10^{14}$~A (Istomin and Sobyanin 2011b). This current is much higher than the characteristic reverse current of positrons causing the polar cap in a normal pulsar to be heated (Beskin et~al. 1993; Barsukov and Tsygan 2003). The specific conditions under which the inner gap is switched on by a lightning are not yet known and require a special study. However, in some cases, the possibility of such a switch-on should be assumed. This is evidenced by the observations of PSR J0941--39 at 1.4~GHz (Burke-Spolaor and Bailes 2010). This source is a kind of a hybrid of intermittent pulsars, nulling pulsars, and RRATs. It is a bright pulsar with a large flux density variation and a quasi-periodic nulling in the on state and behaves as RRAT with a rate of about 90--110 bursts per hour in the off state. It has a triple-humped averaged profile with a width of 105.6~ms, while the width of the brightest recorded single burst is 6~ms. Assuming that $\Delta P\approx105.6$~ms, we obtain $z^{eff}\sim4000$~km using Eq.~\eqref{deltaPsim}, which exceeds $R^{eff}$. If this distance is assumed to be provided by a temporary switch-on of the gap, then a formal lifetime of $\sim13$~ms comparable in order of magnitude to the mentioned width of a single burst will correspond to it. It is the large value of $z^{eff}$ that may suggest the readiness of the given pulsar to switch on after each single burst. This allows the peculiar intermediate position of PSR J0941--39 between normal pulsars and RRATs to be explained. Within the framework of the model being discussed, we can clearly separate the on state when intense plasma generation takes place, as in a typical nulling pulsar, and the off state when no stationary plasma generation takes place, while the separate radio bursts owe their existence to lightnings triggered by the absorption of background gamma-ray photons in the region of open magnetic field lines.

The possibility of the generation of radio emission by an outflowing plasma at great distances is more understandable. The dense relativistic electron-positron plasma with a high multiplicity produced in the lifetime of a separate lightning after a time $T_l$ can outflow from the polar cap regions and can go to distances from the neutron star center that exceed considerably $r_0$ and, hence, can also exceed $R^{eff}$. The radio emission will then be generated at the heights at which the formation of lightnings is inefficient. Hence, the parameter $z^{eff}$ can, in general, exceed the length of the initially formed lightning that enters into Eq.~\eqref{lightningLifetime} to calculate the lifetime. In this sense, the upper limit on $z^{eff}$ in Eq.~\eqref{zEffEstimate} is only a rough estimate.

We will add that a radio pulse propagating through the interstellar medium can be broadened by scattering (Rickett 1977). Shitov et~al. (2009) showed that this pulse broadening for RRAT J2225+35 detected at the Pushchino Radio Astronomy Observatory at a low frequency of 111~MHz is $\tau_{sc}\approx7$~ms for the measured dispersion measure $\text{DM}=51.8\text{ pc\,\,cm}^{-3}$. This is consistent with the statistical dependence of $\tau_{sc}$ on dispersion measure for normal radio pulsars from Kuzmin et~al. (2007), $\tau_{sc}\propto\text{DM}^{2.2\pm0.1}$. The higher the frequency $\nu$ at which the radio pulse is observed, the smaller this broadening, $\tau_{sc}\propto\nu^{-4.1\pm0.3}$. With the exception of RRAT J2225+35, the rotating radio transients are, as a rule, observed at 1.4~GHz.

Thus, based on the available observational data, we cannot unequivocally say what determines the width of the observed radio pulses from RRATs. It can be determined both by the width of the emission cone formed by the lightning emitting regions located at some height above the neutron star surface and by a finite lifetime of the lightning itself. The signal scattering by interstellar irregularities can also give some additional pulse broadening, but its influence depends on the frequency at which the observations are performed.

\section*{THE BURST RATE}

Let
\begin{equation*}
T_l\ll\Delta P\ll\Delta T,
\end{equation*}
i.e., the lightning lifetime is much smaller than the window of radiation, which, in turn, is smaller than the period between two consecutive lightnings. A lightning will then be observed, on average, every $n$ periods, when $n$ can be determined from the condition
\begin{equation}
\label{nEq}
n\Delta P=\Delta T.
\end{equation}
It should be emphasized that here by $n$ we mean precisely the mean number of periods per one recorded burst. The number of periods separating two specific consecutive bursts can be both smaller and larger than this one. The minimum time between two consecutive bursts can be even equal to the rotation period $P$, because the characteristic time of the plasma outflow from the polar cap regions is a factor of $2\pi$ smaller than $P$. This is indicative of a close connection between RRATs and extreme nullers. The corresponding mean burst period is given by the expression
\begin{equation}
\label{burstPeriod}
T=nP=\frac{2\pi}{3}\Delta T\sqrt{\frac{R_L}{z^{eff}}}.
\end{equation}
Formula \eqref{alternativeLightningPeriod} yields the following estimate for the burst period:
\begin{equation}
\label{floorPeriodEstimate}
T\gtrsim\frac{P}{3}\frac{R^{eff}_L}{\sqrt{R_L z^{eff}}}.
\end{equation}

Suppose that $P\sim3$~s, $z^{eff}=R_S$, and $R^{eff}_L=R_L$. Here, we use estimate \eqref{lightningPeriod}. From the relation $R_L/R_S\sim10^4$, we then have
\begin{equation}
\label{TConcreteEstimate}
T\gtrsim100\text{ s}.
\end{equation}
This estimate is typical and corresponds to the formation of radio emission near the neutron star surface. The formation of radio emission not near the surface but at a distance $z^{eff}>R_S$ from the neutron star center leads to a decrease in~$T$. For example, at $z^{eff}\sim R^{eff}\sim100R_S$, the lower limit on the mean period $T$ between two consecutive observed bursts decreases by an order of magnitude, reaching $\sim10$~s.

Another factor reducing $T$ can be the formation of a lightning before the electron-positron plasma produced in a preceding lightning completely escapes from the neutron star magnetosphere (see the remark after Eq.~\eqref{lightningPeriod}). On the other hand, the observational selection effect can lead to an increase in~$T$: not all of the radio bursts can be detected due to a limited sensitivity of radio telescopes. In particular, the existence of this effect is seen from comparison of the results by Esamdin et~al. (2008) and Lyne et~al. (2009). At the Urumqi 25-m radio telescope, 162 radio bursts were recorded from RRAT J1819--1458 at 1.54~GHz in 94~h of observations. In this case, the minimum detectable pulse amplitude is 3.4~Jy at a~$5\sigma$ detection threshold. At the Jodrell Bank 76-m radio telescope, more than 500 bursts were recorded from the same source at 1.4~GHz in 27~h of observations. The pulse rates in these two cases differ by an order of magnitude, being $1.7$ and $19\text{ h}^{-1}$ for the first and second cases, respectively.

Estimate \eqref{TConcreteEstimate} is consistent with the currently available observational data. The mean rate of radio bursts observed at the Parkes 64-m radio telescope at 1.4~GHz for the first 11 discovered rotating radio transients lies within the range from 1~burst in 4~min for J1819--1458 to 1~burst in 3~h for J1911+00 (McLaughlin et~al. 2006). New observational data allow one to talk about the existence of RRATs with a higher burst rate approaching $1\text{ min}^{-1}$ (Keane et~al. 2010). Precisely such data are most interesting, because they provide an independent estimate for the height $z^{eff}$ at which the radio emission is formed. RRAT J1554--52 has a high burst rate, $50.3\text{ h}^{-1}$. The short rotation period, $P\approx0.125$~s, allows the possibility of observing so frequent bursts to be explained even for the formation of emission near the neutron star surface (see~\eqref{floorPeriodEstimate}). The detection of RRAT J1841--14 with a long period, $P\approx6.598$~s, and a high burst rate, $46.0\text{ h}^{-1}$, is more interesting. Assuming that $R^{eff}_L=R_L$, we can obtain an estimate for the effective parameter $z^{eff}\sim25R_S=250$~km using Eq.~\eqref{floorPeriodEstimate}. It is comparable to the lightning formation height $r_0\sim600$~km corresponding to a typical observed burst width of 2~ms \eqref{lightningLifetime}.

Above, we obtained a lower limit on the time interval $T$ between bursts determined by the time $\Delta T$ between two successive lightnings. Obviously, lightnings can be formed more rarely if the gamma-ray background is less intense. The specific value of $\Delta T$ is determined by the photon absorption rate in the region of open magnetic field lines. Let a closed magnetosphere be filled with a dense plasma and the polar cap regions contain no plasma. If intense absorption and scattering of photons is assumed to take place in the closed magnetosphere, then the photons only from a small solid angle $\sim\pi R_P^2/R_S^2\ll1$, where $R_P=R_S^{3/2}/R_L^{1/2}\sim100$~m is the polar cap radius, come into the polar cap region. The conditions under which this occurs are hard to determine, because only the charge density equal to the Goldreich-Julian density $\rho_{GJ}$ is known in a stationary closed neutron star magnetosphere. The electron and positron number densities on which the photon mean free path depends in a stationary closed magnetosphere are not known a priori and can exceed considerably not only the characteristic density $|\rho_{GJ}|/e$ but also the particle density in the polar cap of a normal pulsar. For example, additional continuous ``feeding'' that increases the number of particles but does not change the stationary charge density is provided by continuous absorption of cosmic gamma-ray photons in a closed magnetosphere. Given the above reservations, the photon absorption rate in the region of open magnetic field lines is equal in order of magnitude to
\begin{equation*}
\frac{dN^{abs}_{GB}}{dt}\sim\left(\frac{\pi R_P^{\,2}}{R_S}\right)^2 I^{exp}_{GB}\left(\frac{k^{exp}_{\min}}{k^{eff}_{\min}}\right)^{\beta-1},
\end{equation*}
where $k^{eff}_{\min}=k_{\min}/\chi_{\max}$, $\chi_{\max}\sim0.01$ is the maximum angle between the photon propagation direction and the magnetic field direction in the region of open magnetic field lines (Istomin and Sobyanin 2007). Here, the photons propagate at small angles to the magnetic field direction. Therefore, allowance for the significant deviation of $\sin\chi$ from unity in Eq.~\eqref{weakFieldPhotonAttenuationCoefficient} becomes important in this case. The mentioned formula for $k^{eff}_{\min}$ follows from Eq.~\eqref{kMinValue}, in which the effective magnetic field $B_0^{eff}=B_0\chi_{\max}$ should be substituted for the magnetic field $B_0$ (cf. Eqs.~\eqref{weakFieldPhotonAttenuationCoefficient} and~\eqref{simplifiedAbsorbtionCoefficient}). Assuming that $a_0/\Lambda_\sigma\sim10$, $\chi_{\max}\sim0.01$, and $R_P\sim100$~m and choosing the other parameters to be the same as those used to calculate the flux of photons absorbed in the entire magnetosphere, we have $dN^{abs}_{EGB}/dt\sim0.04\text{ s}^{-1}$. Here, we took into account only the extragalactic photons. Since the diffuse Galactic component gives a flux that is a factor of 2--4 higher (Abdo et~al. 2010), we can write $dN^{abs}_{GB}/dt\sim0.1\text{ s}^{-1}$, which corresponds to $\Delta T\sim10$~s. It can be concluded that no RRATs could exist in the case of an excessively intense cosmic gamma-ray background. Continuous absorption of cosmic gamma-ray photons would cause the polar cap regions of the neutron star magnetosphere to be filled with plasma that would not have time to outflow from the magnetosphere. Consequently, no regions with a high electric field whose existence is necessary for the observed bursty RRAT behavior would have time to be formed.

Consider a neutron star with some fixed rotation period in a cosmic gamma-ray background with a fixed intensity. The photon absorption rate in the region of open magnetic field lines will then increase with surface magnetic field. Formula~\eqref{alternativeLightningPeriod} allows the condition defining the maximum surface magnetic field $B_{\max}$ at which the neutron star is still RRAT to be written:
\begin{equation*}
\label{eqForBMin}
\frac{dN^{abs}_{GB}}{dt}\lesssim\frac{c}{R^{eff}_L}.
\end{equation*}
Note that when this inequality breaks down, the neutron star ceases to be a ``classical'' RRAT due to excessively frequent lightnings and can begin to manifest itself as an extreme nuller. At $R_S\approx10$~km, $R_L\sim10^4 R_S$, $I_{GB}^{exp}\sim(1-5)I_{EGB}^{exp}$, and $R^{eff}_L\sim R^{eff}\sim1000$~km, the maximum surface magnetic field is $B_{\max}\sim(1-4)\times10^{14}$~G. In all RRATs with a known period derivative (McLaughlin et~al. 2009; Lyne et~al. 2009), the surface magnetic field does not exceed $B_{\max}$.

Using Eqs. \eqref{nEq} and \eqref{burstPeriod}, we can independently determine $\Delta T$ directly from observational RRAT timing data. For example, for RRAT J1819--1458 with a period $P\approx4.263$~s, a window of radiation $\Delta P\approx120$~ms, and a time interval between bursts $T\sim180-240$~s (Lyne et~al. 2009), we have $\Delta T\sim5-7$~s. Let us separately consider the already mentioned PSR J0941--39 with a period $P\approx587$~ms and a window of radiation $\Delta P\approx105.6$~ms, which in the off state is RRAT with a time interval between bursts $T\sim33-40$~s (Burke-Spolaor and Bailes 2010). For this source, $\Delta T\sim6-7$~s. The estimates obtained for $\Delta T$ agree well with the above estimate that follows from the photon absorption rate in the region of open magnetic field lines. We see that the time interval $\Delta T$ between two consecutive lightnings can exceed considerably the time of the plasma outflow from the polar cap regions of the neutron star beyond the light cylinder. This provides further evidence for a nonstationary nature of the radio emission from RRATs.

\section*{CONCLUSIONS}

The absorption of a high-energy photon from the cosmic gamma-ray background in a vacuum neutron star magnetosphere gives rise to a lightning. The typical lightning length reaches 1000~km; the lightning radius is 100~m and is comparable to the neutron star polar cap radius. The flux of absorbed photons is large enough for a sufficient number of lightnings to be formed in a short time to fill the inner neutron star magnetosphere with an electron-positron plasma. The magnetosphere filling time is determined not by the flux of absorbed photons but by the lightning formation time. The inner regions of a closed magnetosphere are filled through the absorption of cosmic gamma-ray photons up to distances $R^{eff}\sim1000$~km. The outer magnetosphere can be filled through the transfer of an electron-positron plasma from the inner magnetosphere along magnetic field lines to distances of the order of the light cylinder radius. A necessary condition for the development of a lightning is the existence of a high longitudinal electric field. Such a field can be generated through the plasma outflow from the polar cap regions of the neutron star along open magnetic field lines beyond the light cylinder. If a closed magnetosphere is filled with a dense plasma, then lightnings are efficiently formed only in the region of open magnetic field lines. Each new lightning screens the electric field when it reaches the neutron star surface. The time interval between two consecutive lightnings is limited from below by the time needed for the electron-positron plasma produced by a preceding lightning to outflow from the polar cap regions of the neutron star. The parameters of the generated plasma suggest that the radio emission from separate lightnings can be observed in the form of bright single radio bursts. This emission can be associated with radio bursts from RRATs. To observe the radio emission from a separate lightning, it is necessary that, first, the polar cap of the neutron star be directed toward the observer at some instant of time and, second, the lightning be formed at the same time. The width of single radio bursts can be determined both by the width of the emission cone formed by the lightning emitting regions at some height above the neutron star surface and by a finite lifetime of the lightning itself. The pulse can be additionally broadened by the signal scattering by interstellar irregularities. The radio emission can be formed at great distances exceeding considerably the neutron star radius. The width of the phase distribution for radio bursts from RRATs, along with the width of the averaged profile, is determined by the width of the bundle of open magnetic field lines at the formation height of the radio emission. Although the formation of lightnings is efficient up to the distance $R^{eff}$ from the neutron star center, the radio emission is formed at greater distances in some cases. This suggests that either a short-term switch-on of the inner gap is possible once some of the lightnings have reached the neutron star surface or the electron-positron plasma formed in the relatively short lightning lifetime can generate radio emission already during its outflow along open magnetic field lines. Since the time of the plasma outflow beyond the light cylinder is much shorter than the neutron star rotation period, the time interval between some bursts for a certain gamma-ray background can be equal to the period even without any additional switch-on of the inner gap. The radio burst rate is proportional to the lighting formation rate in the polar cap regions of the neutron star, which, in turn, is determined by the photon absorption rate. However, RRATs could not exist in the case of an excessively intense gamma-ray background. In addition, there exists a maximum surface magnetic field $B_{\max}\sim(1-4)\times10^{14}$~G at which the neutron star can still manifest itself as RRAT. The maximum rate of bursts from RRATs is determined by the time of the plasma outflow from the polar cap region. The typical period of radio bursts from lightnings is $\sim100$~s. The results obtained are consistent with the currently available data and are indicative of a close connection between RRATs, intermittent pulsars, and extreme nullers.

\begin{flushright}
\textit{Translated by V. Astakhov}
\end{flushright}
\end{document}